\documentclass[10pt,aps,prl,twocolumn,superscriptaddress]{revtex4-2}
\usepackage{graphicx}
\usepackage[colorlinks,allcolors=blue]{hyperref}
\usepackage{physics}
\usepackage{bm}
\usepackage{amssymb}
\usepackage[capitalise]{cleveref}

\usepackage{wasysym}

\def\kvec{\mathbf{k}}
\def\qvec{\mathbf{q}}
\def\rvec{\mathbf{r}}
\def\svec{\mathbf{s}}
\newcommand{\Q}{\mathbf{Q}}
\newcommand{\Oc}{\mathcal{O}}

\def\vex#1{\mathbf{#1}}
\def\gvex#1{\boldsymbol{#1}}
\def\bra#1{\mathinner{\langle{#1}|}}
\def\ket#1{\mathinner{|{#1}\rangle}}

\def\braket#1{\mathinner{\langle{#1}\rangle}}

\def\nodagger{{}^{\vphantom{\dagger}}}

\usepackage[dvipsnames,svgnames]{xcolor}

\newcommand{\extra}[1]{\textcolor{cyan}{}}

\begin{document}
	\title{Self-consistent Floquet theory of steady-state ordering in driven open systems}
	\author{Srinjoy Ghosh}
	\affiliation{Department of Physics, Indian Institute of Technology Kanpur,
		Kanpur, Uttar Pradesh 208016, India}
	\author{Babak Seradjeh}
	\affiliation{Department of Physics, Indiana University, Bloomington,
		Indiana 47405, USA}
	\author{Arijit Kundu}
	\affiliation{Department of Physics, Indian Institute of Technology Kanpur,
		Kanpur, Uttar Pradesh 208016, India}
	
	\begin{abstract}
		Periodically driven open systems have quasienergy bands whose steady-state occupations are set by the drive and reservoir. We formulate a self-consistent Floquet spectral theory of electronic ordering in the weak-reservoir-coupling limit, expressing the nonlinear mean-field equation and the particle-hole response in terms of quasienergies, steady-state occupations, and harmonic matrix elements that account for micromotion. We illustrate the theory in the square-lattice Hubbard model and show how the drive produces `dynamical nesting' conditions and how Floquet resonances produce \textit{hysteretic} responses due to multiple self-consistent solutions. Interestingly, we find that the hysteresis branches exhibit distinct Floquet topologies. Away from half-filling, we find that the combination of drive-induced detuning of the van Hove saddle points and the micromotion leads to the onset of switching between competing incommensurate orders.
	\end{abstract}
	\maketitle
	
	\emph{Introduction.---}%
	Periodic driving provides a route to controlling quantum materials beyond equilibrium, with examples ranging from Floquet topological states~\cite{OkaAoki2009,Kitagawa2011,McIver2020} to light-modified superconducting and magnetic responses~\cite{Fausti2011,Mitrano2016,Buzzi2020,Mentink2015,Kennes2018,Walldorf2019,Herre2026,HanSachdev2026}. Instead of energy bands, periodically driven quantum systems are characterized by \textit{quasienergy} bands and periodic Floquet states. While sustained driving in an isolated interacting system generically leads to heating~\cite{LazaridesDasMoessner2014,DAlessioRigol2014,Seetharam2018}, coupling to a reservoir can stabilize a steady state. The steady-state occupations of Floquet states are non-thermal and depend on both the drive and the reservoir~\cite{Genske2015,Seetharam2015,Seetharam2019,Matsyshyn2023,Shi2024,KumariSeradjeh2024}. A theory of electronic ordering must therefore account for both the spectrum and their population.
	
	In equilibrium, band theory makes this connection transparent: the dispersion determines both particle-hole excitation energies and, through the Fermi-Dirac distribution, their occupations. Transition to an ordered state, e.g. due to nesting and Stoner instabilities, can be identified from the band structure.  For driven systems, %Floquet--Keldysh methods, including 
	Floquet dynamical mean-field theory~\cite{TsujiOkaAoki2008,AokiTsuji2014} and the functional renormalization group in Floquet space~\cite{Eissing2016} provide a framework for computing interacting steady states and their response~\cite{OnoIshihara2019}. Recent studies have revealed discontinuous changes of antiferromagnetic order when the drive frequency lies below the equilibrium gap~\cite{Walldorf2019}, hysteretic transitions in light-driven superconductors~\cite{ZhangTakasanTsuji2025}, an AFM--FM reversal at fixed filling~\cite{Herre2026}, and a finite-wavevector magnon-gain instability~\cite{Okugawa2026}. Self-consistent Floquet mean-field and kinetic equations have been used to study symmetry breaking~\cite{Esin2021}, while Keldysh treatments of driven superconductors address the interplay of order, fluctuations, and the bath~\cite{Yang2021,Lin2024}. These studies address different aspects of the ordering problem: the gap equation, competition of different order parameters, and dynamics of fluctuations. 
	
	In this paper, we construct a unified framework for ordering in periodically driven open fermion systems.  Starting from a Floquet-Keldysh action for a driven lattice fermion system coupled to a reservoir, we express the steady-state density matrix, the \emph{nonlinear} mean-field gap equation, and the Gaussian fluctuation kernel in the Floquet eigenbasis. The equal-time Keldysh function becomes diagonal in the Floquet basis when the reservoir linewidth is small compared with the resolved hybridization gaps of the driven spectrum. The resulting ordering problem separates quasienergy nesting, occupation differences, and harmonic vertices, which have no static analog.  
	
	We illustrate the theory in the square-lattice driven Hubbard model. We show that, even when instantaneous dispersion is never nested, Floquet resonances provide a `dynamical nesting' condition for ordering. Moreover, these resonances hybridize the states and redistribute their occupations, resulting in multiple self-consistent solutions and dynamical hysteresis. Interestingly, we find that different hysteretic branches have distinct topologies, as the quasienergy gap closing at resonance changes the Floquet winding numbers. Finally, we show that, away from half-filling, accounting for the drive-induced detuning of the van Hove saddle points and the changes in population and matrix elements due to micromotion, incommensurate orders become competitive. Thus, our formulation connects the discontinuities in antiferromagnetic order~\cite{Walldorf2019} and the AFM--FM reversal~\cite{Herre2026} within a common Floquet spectral theory of ordering. 
	
	The Floquet theory of ordering constructed here applies to any mean-field channel controlled by fermionic susceptibility, opening a new avenue for describing charge, superconducting, and other ordered phases of driven open systems within a common framework.
	
	\emph{Floquet steady-state mean-field theory.}---%
	Consider a fermionic system evolving under a periodically driven single-particle Hamiltonian $H_0(t) = H_0(t+2\pi/\Omega)$ and an undriven interaction, which admits a single-particle mean-field Hamiltonian $H(t) = H_0(t) - m(t) O$ with order parameter $m(t) = I \text{tr}[\rho(t) O]$, where $\rho(t)$ is the density matrix, $I$ is the effective interaction coupling, and $O$ is the ordering operator. The mean-field  quasienergies %$\epsilon_{\vex k\alpha}\in(-\Omega/2,\Omega/2)$, 
	$\epsilon_{\vex k\alpha}$ and Floquet states $\ket{\phi_{\vex k\alpha}(t)} = \ket{\phi_{\vex k\alpha}(t+2\pi/\Omega)}$, where $\vex k$ is the lattice momentum and $\alpha$ labels bands and internal degrees of freedom, are functionals of the order parameter $m(t)$ and satisfy the Floquet-Schr\"odinger equation
	\(
	\left[H(t)-i\partial_t\right]\ket{\phi_{\vex k\alpha}(t)}
	=\epsilon_{\vex k\alpha}\ket{\phi_{\vex k\alpha}(t)}
	\).
	We use natural units $\hbar=k_B=e=1$ unless otherwise noted. Where appropriate, we also set the Bravais lattice spacing $a=1$.
	
	For weak coupling to a wide-band reservoir at temperature $T$ and chemical
	potential $\mu$, the steady-state density operator is diagonal in Floquet states, $\rho(t) = \sum_{\vex k\alpha} f_{\vex k\alpha} \ket{\phi_{\vex k\alpha}(t)}\bra{\phi_{\vex k\alpha}(t)}$ with occupations
	\(
	f_{\kvec\alpha}=\sum_n \braket{\phi^{(n)}_{\kvec\alpha}|\phi^{(n)}_{\kvec\alpha}}
	n_F(\epsilon_{\kvec\alpha}+n\Omega-\mu)
	\)~\cite{Seetharam2015,Matsyshyn2023,Shi2024,Ren2026,KumariSeradjeh2024}.
	Here, the Fermi distribution $n_F(x)=(e^{x/T}+1)^{-1}$, and we have used the Fourier series $\ket{\phi_{\vex k \alpha}(t)} = \sum_{n\in\mathbb{Z}} e^{-in \Omega t} \ket{\phi_{\vex k\alpha}^{(n)}}$ such that the total weight of the Floquet sidebands is normalized $\sum_n \braket{\phi^{(n)}_{\kvec\alpha}|\phi^{(n)}_{\kvec\alpha}} = 1$. Since the steady-state density operator is periodic in time, so is the order parameter, $m(t) = \sum_l e^{-il\Omega t} m^{(l)}$, and the mean-field gap equations for every non-vanishing $m^{(l)}$ take the form
	%%%%
	\begin{equation}
		I^{-1} = \Phi^{(l)}[m] :=\frac{1}{N m^{(l)}}\sum_{\kvec\alpha n}f_{\kvec\alpha}
		\braket{\phi^{(n)}_{\vex k\alpha} | O | \phi^{(n+l)}_{\vex k\alpha}},
		%O^{n,n+l}_{\vex k\alpha,\vex k\alpha}[m]
		\label{eq:gap}
	\end{equation}
	%%%%
	where $N$ is the number of states in the Brillouin zone. %and matrix element $O^{n,n'}_{\vex k\alpha,\vex k'\alpha'} = \braket{\phi^{(n)}_{\vex k\alpha} | O | \phi^{(n')}_{\vex k'\alpha'}}$.
	
	The \textit{cycle-averaged} response to the fluctuations of the order parameter around the mean-field value at wavevector $\vex q$ and frequency $\omega$ is given by the retarded polarization
	%%%%
	\begin{equation}
		\overline{\Pi}^R_{\qvec}(\omega)=\frac{1}{N}\sum_{\substack{\vex k\alpha\beta,l}}
		\frac{(f_{\vex k\alpha}-f_{\vex k+\vex q\beta})
			\left|\sum_n\braket{\phi^{(n)}_{\vex k\alpha} | O | \phi^{(n+l)}_{\vex k+\vex q \beta}}\right|^2}
		{\omega+\epsilon_{\vex k+\vex q \beta}-\epsilon_{\vex k\alpha}+l\Omega+i\gamma},
		\label{eq:pol}
	\end{equation}
	%%%%%
	where $\gamma = 0^+$ is the particle-hole linewidth. The full polarization kernel also encodes the Floquet harmonic response. 
	The susceptibility is $\overline{\chi}^R_{\vex q}(\omega) = \overline{\Pi}^R_{\vex q}(\omega)[I^{-1} - \overline{\Pi}^R_{\vex q}(\omega)]^{-1}$. Thus, the instability of the normal state, signaled by the divergence of the response function $1+\overline{\chi}_{\vex q}^R(\omega\to0) = [1-I\overline{\Pi}^R_{\vex q}(0)]^{-1}$, is found from the Floquet Stoner criterion $I\overline{\Pi}^R_{\qvec}(0) = 1$. See Appendix A for details.
	
	The gap equation~\eqref{eq:gap} and the Lindhard-type polarization function~\eqref{eq:pol} are the main quantities of our approach. They expose the three ingredients of nonequilibrium steady-state ordering under periodic driving: steady-state occupations, Floquet spectrum, and the matrix elements of the ordering operator in the Floquet sidebands.
	
	The driven open systems we study here do not admit a free-energy description, even in the steady state. This is because the occupations $f_{\vex k\alpha}$ depend not only on the quasienergies but also on the micromotion, via the admixture probabilities in the Floquet sidebands. A full discussion on dynamical stability is beyond the scope of the present work.
	
	However, in the following we define a solution to be stable if that can be reached recursively, i.e, the solution has linear stability. The condition for such stability of mean-field solutions is $\overline m\Phi'(\overline m) < 0$ (see Appendix B). In the following whenever we mention `stable' solution, we loosely mean $\overline m\Phi'(\overline m) < 0$, and thus such linear stability and not necessarily dynamical stability.

	\emph{Driven Hubbard model.}---%
	We specialize to the repulsive Hubbard interaction $\frac12 U\sum_i \hat n_i(\hat n_i-1)$, with
	$\hat n_i=\sum_\sigma \hat c^\dagger_{i\sigma} \hat c\nodagger_{i\sigma}$ and $i$ indexing $N$ lattice sites $\vex r_i$. A spatially uniform circularly polarized field enters through
	the Peierls substitution $\vex k\to\vex k+\vex A(t)$, with the vector potential
	$\vex A(t)=A_0(\cos\Omega t,\sin\Omega t)$. Here, the amplitude
	$A_0=\mathcal{E}_0/\Omega$ is dimensionless, where $\mathcal{E}_0$ is the electric field. Decoupling in the SU(2)-symmetric spin channel, using $\hat s_{i\mu}^2 = \hat n_i(2-\hat n_i)$ for each component, yields $I = U/3$, which we use throughout.
	
	\begin{figure}[t]
		\centering
		\includegraphics[width=\linewidth]{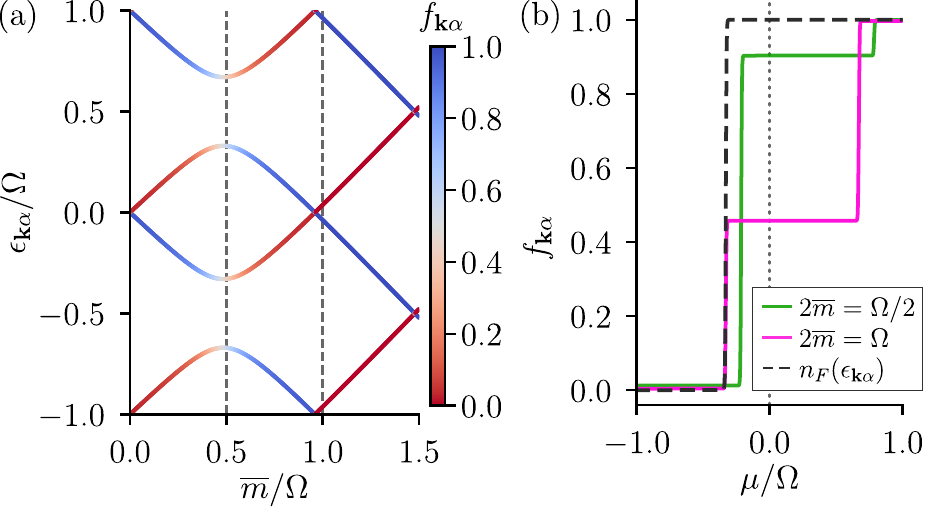}
		\caption{(a) Quasienergies $\epsilon_{\mathbf{k}\alpha}/\Omega$ of the driven mean-field Hamiltonian Eq.~\eqref{eq:sqdriven} at $\mathbf{k} = (\pi/2,\pi/2)$ as a function of the cycle-averaged order parameter $\overline{m}/\Omega$, colored by their steady-state occupation $f_{\mathbf{k}\alpha}$ at reservoir chemical potential $\mu = 0$. The dashed lines mark the Floquet resonance conditions $2\overline{m} = p\Omega$. (b) $f_{\mathbf{k}\alpha}$ of the lower band as a function of the reservoir chemical potential $\mu/\Omega$ for $2\overline{m} = \Omega/2$ (green) and $2\overline{m} = \Omega$ (pink), showing the Fermi--Dirac staircase. The dashed line is the Fermi function $n_F(\epsilon_{\mathbf{k}\alpha}-\mu)$ of the same quasienergy for $2\overline{m} = \Omega$. Drive frequency $\Omega = 4w$ and amplitude $A_0 = 0.5$, with $w'=0$ and temperature $T=0.01w$. 
			%All computations were done with Floquet cutoff $N_{\rm cut}=10$.    
		}
		\label{fig:floq}
	\end{figure}
	
	For the N\'eel order parameter $m_\text{AF}(t) e^{i\vex Q \cdot \vex r_i} \hat{\vex z}$ on the square lattice with $\vex Q=(\pi,\pi)$, the magnetic-zone Hamiltonian acting on the spinor $(\hat c_{\vex k\uparrow},\hat c_{\vex k+\vex Q,\uparrow}, \hat c_{\vex k\downarrow},\hat c_{\vex k+\vex Q,\downarrow})^T$ is
	%%%%
	\begin{equation}
		H_\vex k(t)={\rm diag}[E_{\vex k+\vex A(t)},E_{\vex k + \vex Q+\vex A(t)}]- m_\text{AF}(t)\sigma_3\tau_1,
		\label{eq:sqdriven}
	\end{equation}
	%%%%
	where $E_{\vex k}=-2w\sum_{\mu}\cos k_\mu - 4w' \prod_\mu \cos k_\mu$ with $w$ and $w'$ the nearest- and next-nearest-neighbor hopping, respectively, and Pauli matrices  $\gvex\tau$ act on the $(\vex k,\vex k+\vex Q)$ block. Further details are in Appendix C.
	
	\emph{Dynamical nesting.}---%
	The Floquet Stoner criterion for the onset of ordering may be satisfied when there is a nesting condition relating the quasienergies in the denominator of \cref{eq:pol} as $\epsilon_{\vex k\alpha}-\epsilon_{\vex k+\vex Q\beta}=l\Omega$ along an extended momentum manifold. When this nesting condition holds with a nonzero matrix element of the ordering operator, integration over the nested manifold gives a logarithmic enhancement of the polarization weighted by the bath occupations and cut off by temperature and linewidth. For the static model on the square lattice with $w'=0$ at half filling, this results in antiferromagnetic ordering at an arbitrarily small interaction as temperature $T\to0$. This nesting is lifted away from half-filling or for $w'\neq0$.
	
	In the normal state, $m_{\mathrm AF} = 0$, the Hamiltonian commutes with itself at all times. The one-period evolution operator is then $U_{\vex k}(2\pi/\Omega,0)=e^{-i H^{\rm eff}_{\vex k}2\pi/\Omega}$, with the exact effective Hamiltonian $H^{\rm eff}_{\vex k} = {\rm diag}[\overline{E}_{\vex k},\overline{E}_{\vex k + \vex Q}]$, where $\overline E_{\vex k}$ is the cycle-averaged dispersion with the nearest- and next-nearest-neighbor hoppings renormalized to $wJ_0(A_0)$ and $w'J_0(\sqrt2A_0)$, respectively. As $A_0$ is increased, the latter vanishes first at $A_0\equiv A_*\simeq1.70$, and the Floquet bands become nested with $l=0$, although the instantaneous bands are not nested. We have checked numerically that the susceptibility indeed diverges (Appendix C). For $m_{\mathrm AF} \neq 0$ the Hamiltonian no longer commutes with itself at different times and the cycle average $H^{\rm eff}_{\vex k} = {\rm diag}[\overline{E}_{\vex k},\overline{E}_{\vex k + \vex Q}] - \overline m \,\sigma_3 \tau_1$ is the leading term of the high-frequency expansion of the Hamiltonian, with $\overline m \equiv m_\text{AF}^{(0)}$, which keeps the renormalized bands but neglects micromotion effects.

	%\babak{Replace $m_0$ with $\overline m$. What is $\mu$ in (a)? The red/blue color in (b) is confused with the color legend in (a), maybe use green / orange (???) or change the color map in (a) to grayscale? Remove the $A_0$ and $\Omega$ values from the panels; these are stated in the caption and make the figure too busy. Also, maybe one of the vertical axis labels, $f_{\vex k\alpha}$, can be removed, and the other repositioned to serve as the label for both panels? That would be cleaner.}
	
	\begin{figure}[t]
		\centering
		\includegraphics[width=.9\linewidth]{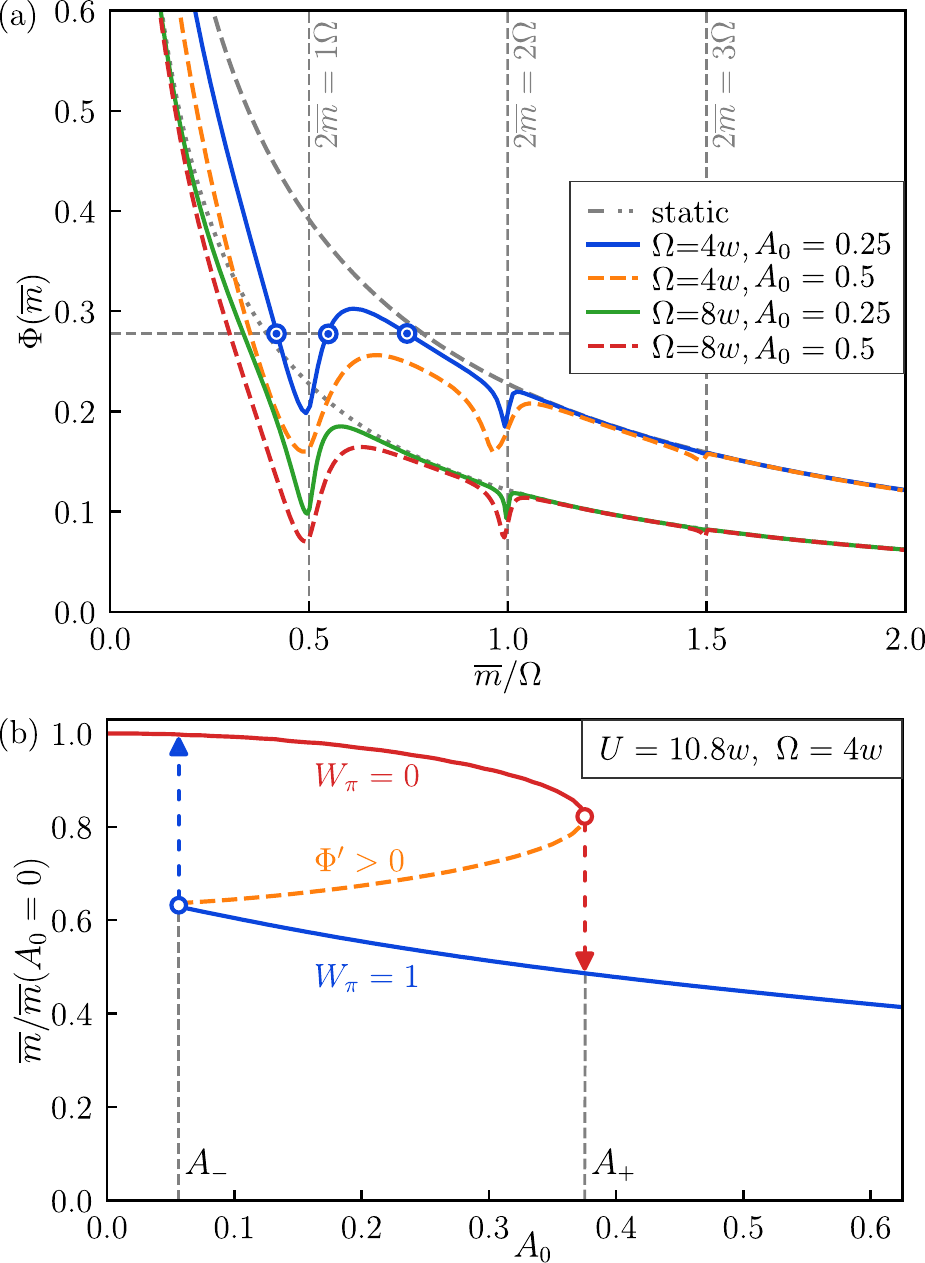}
		\caption{(a) $\Phi(\overline{m})$, Eq.~\eqref{eq:gap}, as a function of the cycle-averaged N\'eel amplitude $\overline{m} / \Omega$ for the nearest-neighbor square lattice for frequency $\Omega = 4w$ and $8w$, each with amplitude $A_0 = 0.25$ and $0.5$. The grey curves are the undriven $\Phi$ for the two values of $\Omega$. The horizontal dashed line is {at $3 / U$} at $U = 10.8w$, and its three intersections with the $\Omega=4w$, $A_0 = 0.25$ curve are circled. (b) $\overline{m}$, normalized by its undriven value, as a function of $A_0$ at $\Omega = 4w$ and $U = 10.8w$. The outer branches, with $\Phi' < 0$, carry the $\pi$-gap winding numbers $W_\pi = 0$ (red) and $W_\pi = 1$ (blue). The dashed middle branch has $\Phi' > 0$ and is unstable. Arrows mark the jumps at the folds $A_+$ and $A_-$ under increasing and decreasing drive, respectively. We use $200\times200$ momentum grid and temperature $T = 0.01w$.}
		\label{fig:meanf}
	\end{figure}
	
	\emph{Floquet resonances and dynamical hysteresis.}---%
	A remarkable effect of the interplay of the Floquet spectrum and steady-state occupation is dynamical hysteresis manifested at Floquet resonances. To see this, we first approximate $m_\text{AF}(t) \approx m_\text{AF}^{(0)} \equiv \overline m$. For the square lattice, $E_{\vex k + \vex A(t)} = \sum_l e^{-il\Omega t} E_{\vex k}^{(l)}$, we separate $E^{(0)}_{\vex k} \equiv \overline E_{\vex k}$ and treat the higher harmonics $E^{(l\neq0)}_{\vex k}$ perturbatively. Then, the quasienergies bands are $\pm [{(\overline E_{\vex k})^2+\overline m^2}]^{1/2}$, hybridized by $E^{(l\neq0)}_{\vex k}$ whenever the corresponding matrix element is nonzero. The hybridization opens avoided crossings at Floquet resonances
	%%%%
	\begin{equation}
		\Delta_{\vex k}(\overline m) \equiv 2[{(\overline E_{\vex k})^2+\overline m^2}]^{1/2} = l\Omega,
		\label{eq:res}
	\end{equation}  
	%%%%
	through virtual coupling to nonresonant replicas at first order in $E^{(l\neq 0)}_{\vex k}/\Omega$. On the nested surface, which is the boundary of the magnetic Brillouin zone, $\overline E_{\vex k} = 0$ and the band separation is $2\overline m$, yielding Floquet resonances near every $2\overline m= p\Omega$ with integer $p$.
	
	The steady-state occupation of the hybridized resonant states converges to one-half at equal mixing. \Cref{fig:floq}(a) shows the evolution of the Floquet bands and their steady-state occupation with $\overline m$ for $w'=0$ at $\mu=0$. An avoided crossing is obtained near $\overline m = \Omega/2$ corresponding to $l=1$ Floquet resonance. The dependence on $\mu$ is shown in \cref{fig:floq}(b): each harmonic of a Floquet state samples the reservoir in a different Floquet zone, producing a staircase rather than a single Fermi step. At $\mu=0$, the lower band is almost fully occupied away from resonance, but its occupation approaches one-half near the $l=1$ resonance. 
	
	%\textcolor{red}{just a note: at the momentum of 1(a), there is a crossing, at other momentum there is avoided crossing, thats why they show up at 2(a).}
	
	The near equality of the occupations of resonant Floquet states suppresses their contribution to ordering. We can see this from the suppression in \cref{eq:pol}. To see this more explicitly, we plot $\Phi(\overline m)$ in~\cref{fig:meanf}(a). In the static model, $\Phi(\overline m)$ decreases monotonically, so there is at most one positive solution for $\overline m$ at a given interaction. Under driving, the suppression of contributions of Floquet resonant states produces notches in $\Phi(\overline m)$. For $I^{-1}$ near the values of the notch in $\Phi(\overline m)$, there is an odd number of solutions to the mean-field gap equation~\cref{eq:gap}. However, solutions with $\Phi'(\overline m) > 0$ are unstable.

	The number of solutions changes as drive parameters such as amplitude and frequency are varied. In \cref{fig:meanf}(b), we vary the amplitude $A_0$ at fixed frequency. At weak drive, there is a single solution $\overline m_1$, close to the undriven value. Increasing $A_0$ lowers $\Phi(\overline m)$, most strongly near the resonance $\overline m = \Omega/2$. At $A_-$, the minimum of $\Phi$ near the resonance reaches $I^{-1}$, and a second stable solution $\overline m_2$ appears, together with an unstable one between the two. At $A_+$, the local maximum of $\Phi$ at $\overline m > \Omega/2$ falls below $I^{-1}$, and $\overline m_1$ disappears together with the unstable solution. Thus, by slowly increasing the amplitude, the order parameter follows $\overline m_1$ up to $A_+$ and then jumps to $\overline m_2$. Conversely, by slowly decreasing the amplitude, it follows $\overline m_2$ down to $A_-$ and then jumps to $\overline m_1$. The order parameter therefore traces a hysteresis loop: between the folds $A_-$ and $A_+$, its value depends on the history of the drive.
	
	Therefore, Floquet resonances explain the hysteresis observed in numerical studies~\cite{Walldorf2019,ZhangTakasanTsuji2025}. For example, the $l=1$ notch accounts both for the discontinuities reported in Ref.~\cite{Walldorf2019} and for their disappearance when $\Omega$ exceeds the equilibrium gap.

	\emph{Nontrivial Floquet topology.}---%
	Since quasienergy gaps close at Floquet resonances, the topology of the Floquet bands may also change. We can characterize this Floquet topology using the winding numbers $W_0$ and $W_\pi$ of the periodicized one-period evolution operator 
	for quasi energy gaps at $0$ and $\Omega/2$, respectively, and band Chern number $C=W_0-W_\pi$~\cite{Rudner2013,Usaj2014,Wang2016}.
	In our model~\eqref{eq:sqdriven}, the driven normal state is topologically trivial. At finite $\overline{m}$, the avoided crossing at a Floquet resonance, \cref{eq:res}, lies at quasienergy $l\Omega/2$, so even values of $l$ may change $W_0$ while odd values of $l$ may change $W_\pi$.
	
	In the range of parameters used in \cref{fig:meanf}(b), the Floquet resonance occurs for $l=1$. We show the calculated values of $W_\pi$ for the two stable solutions $\overline m_1 > \overline m_2$, finding $W_\pi(\overline m_1) = 0$ and $W_\pi(\overline m_2) = 1$ (per spin), with identical $W_0$ in either branch. Thus, the branches are topologically different and are predicted to have signatures in terms of chiral edge modes.
	
	%\color{black}
	\begin{figure}[t]
		\centering
		\includegraphics[width=\linewidth]{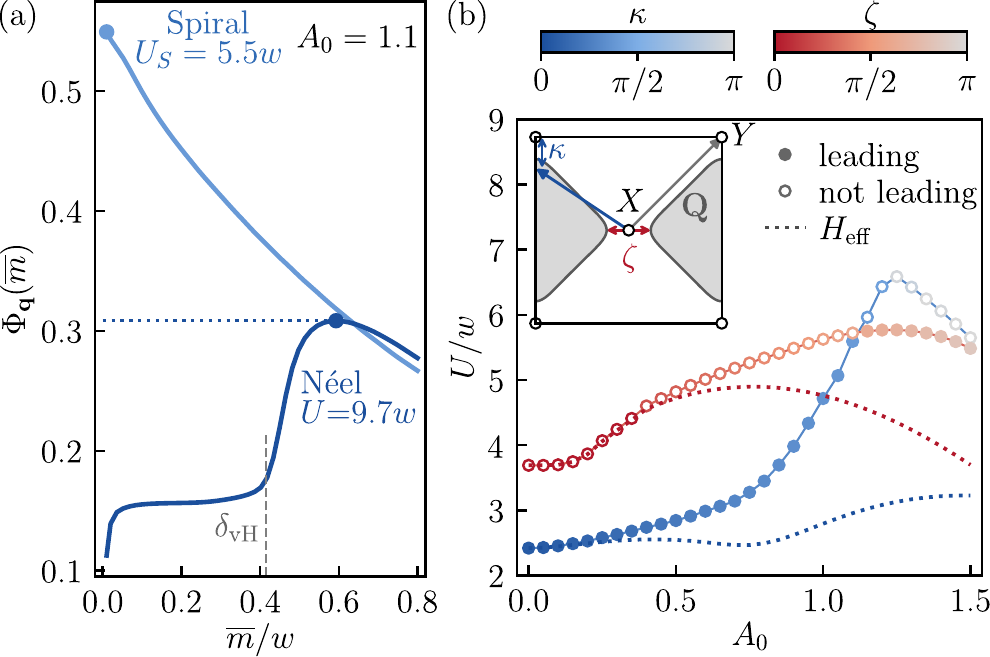}
		\caption{(a) $\Phi_{\mathbf q}(\overline m)$ at $A_0=1.1$ for N\'eel order ($\mathbf q=\mathbf Q$) and for the spiral at its onset wavevector, colored by $\kappa$ as in (b). Dots mark the maxima, where each order first appears. The dotted line is $3/U$ at the N\'eel onset and the dashed line marks $\overline m = \delta_{\rm vH}$. (b) Onset interaction of the spiral $(\pi, \pi-\kappa)$ and axial $(\zeta,0)$ families versus $A_0$. $\kappa = 0$ denotes N\'eel order and $\zeta=0$ denotes uniform ferromagnetic order - the two families meet at $X=(\pi, 0)$ where $\kappa = \zeta = \pi$. Filled circles mark the leading instability of the normal state. The dotted lines use $H_{\rm eff}$ with Fermi occupations. Inset: the zone at $A_0=1.1$ ($0 < k_x < 2\pi$) with filled states shaded, showing $\mathbf Q$, the spiral wavevector and the axial pair $(\zeta,0)$ drawn from the empty saddles $X$ and $Y$. The parameters are $w'=-0.2w$, $\mu=4w'$, $\Omega=4w$, $T=0.01w$.  We use a $380\times380$ momentum grid.
		}
		\label{fig:vh}
	\end{figure}
	
	\emph{Competing orders near van Hove saddle points.}---%
	Away from dynamical nesting at $A_*$, or from $\mu=0$, the N\'eel and other orders can compete. The competition is sharpest when $\mu$ lies at a saddle point. In our model~\eqref{eq:sqdriven}, the van Hove saddle sits at $4w'$ without the drive~\cite{LinHirsch1987,Hlubina1997,Alvarez1998,HonerkampSalmhofer2001,Igoshev2010}. The saddle points $X=(\pi,0)$ and $Y=(0,\pi)$ enhance both the N\'eel order, which pairs $X$ with $Y$~\cite{RiceScott1975}, and uniform order, which pairs states within one saddle. To study these competing orders, we take $w'=-0.2w$ and fix the reservoir chemical potential at $\mu=4w'$. The drive raises the saddle point a distance $\delta_{\rm vH}\equiv4|w'[1-J_0(\sqrt2A_0)]|$ above $\mu$.
	
	We study competing orders at wavevector $\vex q$, coupling $\hat c_{\vex k \uparrow}$ to $\hat c_{\vex k+\vex q\downarrow}$ using the Hamiltonian $H_{\vex k, \vex q}(t) = \mathrm{diag}[E_{\kvec+\vex A(t)},E_{\kvec+\qvec+\vex A(t)}]-m(t)\tau_1$. The N\'eel and uniform order correspond to $\qvec=\Q$ and $\qvec=0$, respectively. Keeping only the time-averaged order parameter, we define $\Phi_\qvec(\overline m)$ by \cref{eq:gap}, evaluated in the Floquet states of $H_{\kvec,\qvec}$ with $O=\tau_1$ and $I=U/3$. Using the Floquet Stoner criterion, the order sets in for $U > U_{\rm S}(\qvec)=3/\max_{\overline m}\Phi_\qvec(\overline m)$. We find $\Phi_{\vex q}(\overline m)$ has its maximum at finite $\overline m$ for both the N\'eel and uniform order, hence their onset is discontinuous. This is because the %drive lifts the saddle point. The 
	N\'eel order splits the states at $X$ and $Y$ saddles to $\mu+\delta_{\rm vH}\pm\overline m$. They contribute appreciably to $\Phi_\Q$ only once $\overline m>\delta_{\rm vH}$, when the lower level falls below $\mu$. Thus, $\Phi_\Q$ rises steeply near $\delta_{\rm vH}$ and peaks at finite $\overline m$ [see~\cref{fig:vh}(a)]. The same argument applies to uniform order, which must lower the saddles of one spin by $\delta_{\rm vH}$ down to $\mu$.
	
	These discontinuous transitions may be avoided by the continuous onset of an incommensurate order. A similar phenomenology leads to finite-momentum pairing for a superconductor in an exchange field~\cite{Sarma1963,FuldeFerrell1964,LarkinOvchinnikov1964} and for an imbalanced exciton condensate~\cite{Seradjeh_2012}, and to incommensurate order for an imperfectly nested antiferromagnet~\cite{Rice1970,Vorontsov2010}.
	
	To test this, we also consider a \textit{spiral} order with $\vex q = (\pi,\pi-\kappa)$ and an \textit{axial} order with $\vex q = (\zeta,0)$. The spiral order mixes the $X$ saddle with $(0,\pi-\kappa)$ near the $Y$ saddle, which lies below $\mu$ once $\kappa$ is large enough. One level of each such pair is then filled at any $\overline m$, and the spiral can set in continuously [see~\cref{fig:vh}(a)]. The axial order does the same within one saddle, mixing the two states at $\mu$ on either side of it. %The spiral thus extends N\'eel order, and the axial order extends uniform order [\cref{fig:vh}(b), inset].
	
	In~\cref{fig:vh}(b), we show the solution of each incommensurate order with the lowest interaction threshold. For $A_0\lesssim1.1$, we find the spiral order sets in at a lower threshold than the axial or the commensurate orders. The axial order overtakes at larger driving amplitude. These are the leading static instabilities of the normal state and are stable against admixing with other orders (see Appendix B).
	
	While these explanations rely only on the quasienergy bands, our solutions of the Floquet gap equation include the effects of the micromotion, both through the steady-state occupations and the matrix elements in the Floquet basis. We also compute the mean-field order parameter from the effective Hamiltonian $H^{\rm eff}_{\vex k}$ with equilibrium Fermi distribution, which we show by dotted curves in~\cref{fig:vh}(b). The latter show no switching from spiral to axial order for $A_0<1.5$. This is consistent with	AFM--FM reversal at fixed filling found in Ref.~\cite{Herre2026}. At fixed filling, as saddle points move, the Fermi level also drops. This pushes the discontinuous onsets of the commensurate orders to larger $U$. Details of fixed filling study is presented in Appendix D.
	
	\emph{Concluding remarks.}---% 
	We have formulated a unified mean-field Floquet theory of steady-state ordering, based on the functional $\Phi[m]$, \cref{eq:gap}, of a time-dependent order parameter $m(t)$. We illustrated this framework in the driven Hubbard model on the square lattice, finding dynamical nesting and topologically nontrivial hysteresis of commensurate order due to Floquet resonances at half filling, and competing incommensurate ordering away from half filling, due to detuned saddle points and micromotion.

	Floquet-Bloch states have been resolved on topological-insulator surfaces and in graphene~\cite{Wang2013,Merboldt2025,Choi2025}, followed on subcycle timescales~\cite{Ito2023}, and approached in steady state under continuous microwave driving~\cite{Park2022}. Our findings can be tested both in spectral and transport studies.
	The possibility of heterodyne ordering with $l\neq0$ and the dynamical stability, including possible magnon gain~\cite{Okugawa2026} of mean-field solutions, are interesting problems for future studies.

	\begin{acknowledgments}
		This work is supported in part by the National Science Foundation (USA) through Grant No. DMR-2533543 (B.S.) and A.K. acknowledges funding support from ANRF (India) via Grants No. ANRF/ARG/2025/002460/PS and No. ANRF/ARGM/2025/002682/TS.
	\end{acknowledgments}
	
	\bibliography{refs}

	% ---------- appendix: replaces \appendix ----------
	\clearpage
	\begin{center}\textbf{Appendices}\end{center}
	\pdfbookmark[1]{Appendix}{appendix}
	\setcounter{subsection}{0}
	\setcounter{equation}{0}
	\setcounter{figure}{0}
	\setcounter{table}{0}
	\setcounter{secnumdepth}{2}          % letter the parts; subsubsections stay unnumbered
	\renewcommand{\thesubsection}{\Alph{subsection}}
	\renewcommand{\theequation}{S.\arabic{equation}}
	\renewcommand{\thefigure}{S.\arabic{figure}}
	\renewcommand{\thetable}{S.\arabic{table}}
	\renewcommand{\theHequation}{S.\arabic{equation}}   % keeps hyperlinks pointing to the right place
	\renewcommand{\theHfigure}{S.\arabic{figure}}
	\renewcommand{\theHtable}{S.\arabic{table}}

	\subsection{Floquet theory and Green's functions}
	
	\subsubsection{Floquet theory}
	
	We consider a time-periodic single-particle Hamiltonian $H_0(t)=H_0(t+T)$ where $T$ is the drive period and $\Omega = 2\pi/T$ is the drive frequency. Floquet's theorem states that the solutions of the time-dependent Schrödinger equation with a time-periodic Hamiltonian can be written as 
	\begin{equation*}
		|\psi_\alpha(t)\rangle = e^{-i\epsilon_\alpha t} |\phi_\alpha(t)\rangle,
	\end{equation*}
	where $\epsilon_\alpha$ is the quasienergy and $|\phi_\alpha (t)\rangle = |\phi_\alpha (t + T)\rangle $ is the time-periodic Floquet state. We can decompose the Floquet state into Fourier modes $|\phi_\alpha(t)\rangle = \sum_{n\in\mathbb{Z}}e^{-i n\Omega t}|\phi^{(n)}_\alpha\rangle$, and the eigenvalue problem written in terms of these Fourier modes is a time-independent Schrödinger equation given by
	\begin{equation*}
		\sum_m\left( H_{n-m}-n\Omega\delta_{nm}\mathbb{I} \right)|\phi^{(m)}_\alpha\rangle	=\epsilon_\alpha |
		\phi^{(n)}_\alpha\rangle ,
	\end{equation*}
	with $H_n = \frac{1}{T} \int_{0}^{T} dt  e^{in\Omega t} H_0(t)$ and $\epsilon_\alpha \in [-\Omega / 2, \Omega / 2)$. In this so-called Floquet space, the completeness condition for the Fourier modes is given by
	\begin{equation*}
		\sum_{\alpha,r} |\phi^{(n+r)}_\alpha\rangle \langle\phi^{(m+r)}_\alpha|
		=\delta_{nm}\mathbb{I}.
	\end{equation*}
	
	\subsubsection{Keldysh Green's functions} 
	
	We now apply Floquet theory to non-equilibrium Green's functions defined on the Keldysh contour. A periodically driven system is said to have reached a non-equilibrium steady state when its Green's functions are time-periodic in both indices, that is, $G(t+T,t'+T) = G(t,t')$. For any function satisfying such a condition, we define a two-index Floquet representation \cite{TsujiOkaAoki2008}
	\begin{equation*}
		G_{mn}(\omega)=\frac1T\int_0^T\!\!dt_1\int_{-\infty}^{\infty}\!\!dt_2\;
		e^{i(\omega+m\Omega)t_1}e^{-i(\omega+n\Omega)t_2} G(t_1,t_2),
	\end{equation*}
	with $\omega\in[-\Omega/2,\Omega/2)$.  A time convolution  $A = B \circ C$ is then converted to a matrix product in Floquet space,
	\begin{equation*}
		[A\circ B]_{mn}(\omega)=\sum_l A_{ml}(\omega)B_{ln}(\omega).
	\end{equation*}
	For a free fermionic Hamiltonian $H(t) = \sum_k \hat{\psi}_k^\dagger H_0(t) \hat{\psi}_k$ we can define the retarded, advanced and Keldysh Green's functions, which can be written in the Floquet representation as
	\begin{align} \label{eq: suppgreen}
		G^R_{mn}(\omega) & = \sum_{\alpha,l}
		\frac{|\phi^{(m+l)}_\alpha\rangle \langle\phi^{(n+l)}_\alpha|}
		{\omega-\epsilon_\alpha-l\Omega + i0^+},\nonumber \\
		G^A_{mn}(\omega) &= \left[G^R_{nm}(\omega)\right]^\dagger
		=\sum_{\alpha,l}
		\frac{|\phi^{(m+l)}_\alpha\rangle \langle\phi^{(n+l)}_\alpha|}
		{\omega-\epsilon_\alpha-l\Omega - i0^+}, \\
		G^K_{mn}(\omega) & = -2\pi i \sum_{\alpha, l} (1 - 2f_{\alpha}) \delta(\omega - \epsilon_\alpha - l\Omega) |\phi^{(m+l)}_\alpha\rangle \langle\phi^{(n+l)}_\alpha| \nonumber
	\end{align}
	where $f_\alpha$ is the occupation of the Floquet states. 
	
	\subsubsection{Driven-dissipative system}
	
	We now couple the system locally to a non-interacting fermionic reservoir,
	\begin{equation} \label{eq: supphres}
		H_{\rm res}=\sum_{\lambda,p}\hat\phi^{\lambda\dagger}_p H_\lambda\hat\phi^\lambda_p
		+\sum_{\lambda,k,p}\big(\hat\psi^\dagger_k V_\lambda\hat\phi^\lambda_p+{\rm h.c.}\big).
	\end{equation}
	where $\lambda$ is the reservoir index. We can integrate out the reservoir fields, getting the following self-energy terms in the wide-band limit:
	\begin{equation*}
		\Sigma^{R/A}_{mn}(\omega) = \mp i\Gamma\delta_{mn},
		~~
		\Sigma^K_{mn}(\omega) = -i \sum_{\lambda} \Gamma_\lambda F_\lambda(\omega + n\Omega) \delta_{mn}.
	\end{equation*}
	where $F_\lambda(x)=1-2n^\lambda_F(x-\mu_\lambda)=\tanh(\frac{x-\mu_\lambda}{2T_\lambda})$ is the bath distribution function and $\Gamma = 1/2 \sum_{\lambda} \Gamma_\lambda = \sum_{\lambda} \pi |V_\lambda|^2 \rho_\lambda$ is the reservoir broadening. Eqs.~\eqref{eq: suppgreen} can now be rewritten as
	\begin{align*} 
		& G^R_{mn}(\omega) = \sum_{\alpha,l} \frac{|\phi^{(m+l)-}_\alpha\rangle \langle\phi^{(n+l)+}_\alpha|} {\omega-\epsilon_\alpha-l\Omega + i\gamma_\alpha}, \\
		&  G^A_{mn}(\omega) = \sum_{\alpha,l} \frac{|\phi^{(m+l)+}_\alpha\rangle \langle\phi^{(n+l)-}_\alpha|}	{\omega-\epsilon_\alpha-l\Omega - i\gamma_\alpha}, \nonumber 
	\end{align*}
	where we define 
	\begin{align*}
		(H_0(t) \pm i\Gamma - i\partial_t) |\phi_\alpha^\pm\rangle = (\epsilon_\alpha \pm i\gamma_\alpha) |\phi_\alpha^\pm\rangle.
	\end{align*}
	These eigenkets satisfy the normalization condition $\langle\phi_\alpha^+(t)|\phi_\beta^-(t)\rangle = \delta_{\alpha\beta}$, and the completeness relation $\sum_\alpha |\phi_\alpha^-(t)\rangle\langle\phi_\alpha^+(t)| = \mathbb{I}$.
	In the steady-state, where initial memory effects are not retained, the Keldysh Green's function is now given by the Keldysh-Dyson equation
	\begin{equation} \label{eq: suppdyson}
		G^K=G^R \circ \Sigma^K \circ G^A.
	\end{equation}
	Substituting the Green's functions into Eq.~\eqref{eq: suppdyson}, we get the steady-state Keldysh Green's function,
	\begin{align*}
		G^K_{mn}(\omega) = &-i \sum_{\alpha, \beta} \sum_{l, l'} |\phi_\alpha^{(m+l)-}\rangle \langle\phi_\beta^{(n+l')-}| \\
		&\times\sum_{n', \lambda} \frac{\langle\phi_\alpha^{(n'+l)+}| \Gamma_\lambda |\phi_\beta^{(n'+l')+}\rangle F_\lambda(\omega + n'\Omega)}{(\omega - \epsilon_\alpha - l\Omega + i\gamma_\alpha)(\omega - \epsilon_\beta - l'\Omega - i\gamma_\beta)}.
	\end{align*}
	We write the equal-time Keldysh Green's function,
	\begin{widetext}
		\begin{align*}
			G^K(t,t) &= -i \sum_{\lambda} \sum_{\alpha, \beta} \sum_{p, l, l'} \int \frac{d\omega}{2\pi} \frac{e^{i(l - l')\Omega t}\langle\phi_\alpha^{(p+l)+}| \Gamma_\lambda |\phi_\beta^{(p+l')+}\rangle F_\lambda(\omega + p\Omega) |\phi_\alpha^{-}(t)\rangle \langle\phi_\beta^{-}(t)|}{(\omega - \epsilon_\alpha - l\Omega + i\gamma_\alpha)(\omega - \epsilon_\beta - l'\Omega - i\gamma_\beta)}.
		\end{align*}
	\end{widetext}
	We compute the integral over $\omega$ by expanding the Fermi function and using Matsubara summation techniques. In the weak coupling limit $\Gamma^\lambda \to 0$, the non-vanishing terms occur for $\alpha = \beta$ and $l = l'$. Keeping terms only upto leading order in $\Gamma^\lambda$, we get
	\begin{align*}
		G^K(t,t) &= \sum_{\alpha} |\phi_\alpha(t)\rangle \langle\phi_\alpha(t)| \\ 
		& \times\sum_{\lambda, n} \frac{\langle \phi_\alpha^{(n)}| \Gamma^\lambda |\phi_\alpha^{(n)}\rangle}{2i\gamma_\alpha} \left( 1 - 2n_F^\lambda(\epsilon_\alpha + n\Omega - \mu_\lambda) \right) .
	\end{align*}
	For a thermal bath connected with the same tunnelling strength, the Keldysh function becomes 
	\begin{equation*}
		G^K(t,t) = -i \sum_{\alpha} (1 - 2f_\alpha) |\phi_\alpha(t)\rangle \langle\phi_\alpha(t)|, 
	\end{equation*}
	where 
	\begin{equation} \label{eq: suppqoc}
		f_\alpha = \sum_{n} \langle\phi^{(n)}_\alpha|\phi^{(n)}_\alpha\rangle n_F(\epsilon_\alpha + n\Omega - \mu),
	\end{equation}
	is the steady-state quasienergy occupation. Using $G^<=\tfrac12(G^K-G^R+G^A)$ with $G^R(t,t)-G^A(t,t)=-i\mathbb{I}$, we get the steady-state density matrix for the system,
	\begin{equation} \label{eq: supprho}
		\rho(t)=-iG^<(t,t)=\sum_\alpha f_\alpha|\phi_\alpha(t)\rangle\langle\phi_\alpha(t)| .
	\end{equation}
	
	\subsubsection{Ordering and the saddle-point equation} \label{sec:saddle}
	
	\noindent Consider an interaction decoupled in a fermion-bilinear channel
	\begin{equation*}
		\hat O_a(\qvec)=\sum_\kvec\hat\psi^\dagger_{\kvec+\qvec}\mathcal{O}_a\hat\psi_\kvec
	\end{equation*}
	with coupling $I_a$.  On the Keldysh contour the Hubbard--Stratonovich weight is $-m_a^2/2I_a$ at each contour time - we rotate the fields on the contour as
	\begin{equation*}
		m^c=\tfrac12(m^++m^-),~~ m^q=\tfrac12(m^+-m^-),
	\end{equation*}
	so that the bare term becomes $-(2N/I_a)\int dt\sum_\kvec m^c_{a,\kvec}m^q_{a,-\kvec}$, and we rotate the fermions with the Larkin--Ovchinnikov convention
	\begin{align*}
		&	\psi^{1,2}=\tfrac{1}{\sqrt2}(\psi^+\pm\psi^-),\\
		&	\bar\psi^1=\tfrac{1}{\sqrt2}(\bar\psi^+-\bar\psi^-),\\
		&	\bar\psi^2=\tfrac{1}{\sqrt2}(\bar\psi^++\bar\psi^-),
	\end{align*}
	for which the vertex is $\hat m=m^c\hat\tau_0+m^q\hat\tau_1$. Integrating out the fermions gives an effective action $S_{\rm eff}[m^c,m^q]$, and the saddle condition
	$\delta S_{\rm eff}/\delta m^q_a=0$ at $m^q=0$, evaluated with Eq.~\eqref{eq: supprho} yields
	\begin{equation} \label{eq: suppsaddle}
		m^{(l)}_a=\frac{I_a}{N}\sum_{\kvec,\alpha}f_{\kvec\alpha}\sum_n
		\langle \phi^{(n)}_{\kvec\alpha}|{\mathcal{O}_a}|\phi^{(n+l)}_{\kvec\alpha}\rangle .
	\end{equation}
	where the Floquet states (and consequently, the quasienergy occupations) have an additional momentum index. The quasienergies, occupations and Floquet states in Eq.~\eqref{eq: suppsaddle} are evaluated in the self-consistent `mean-field' Hamiltonian containing the complete set of $m^{(l)}_a$.
	
	\subsubsection{Gaussian action of fluctuations} 
	Expanding the fermionic effective action to quadratic order around a steady-state saddle point gives \cite{KamenevBook2023}
	\begin{align}  \label{eq: suppkel}
		S^{(2)}=-N\int\!\frac{d\omega}{2\pi}&\sum_\qvec
		\begin{pmatrix}\delta m^c_{-\qvec}&\delta m^q_{-\qvec}\end{pmatrix}\\
		&\times\begin{pmatrix}
			0 & I^{-1}-\Pi^A_\qvec(\omega)\\[2pt]
			I^{-1}-\Pi^R_\qvec(\omega) & -\Pi^K_\qvec(\omega)
		\end{pmatrix}
		\begin{pmatrix}\delta m^c_{\qvec}\\[2pt] \delta m^q_{\qvec}\end{pmatrix}. \nonumber
	\end{align}
	Each block is a matrix in ordering channels, sublattice, and external Floquet harmonics. At fixed quasienergy occupations, the retarded polarizability is
	\begin{align} \label{eq: supppol}
		\Pi^{R,ab}_{\mathbf q,mn}(\omega)&=\frac1N\sum_{\mathbf k, \alpha\beta,l}\frac{(f_{\mathbf k\alpha}-f_{\mathbf k+\mathbf q,\beta})U^{a,lm}_{\beta\alpha}(\mathbf k,\mathbf q)V^{b,nl}_{\alpha\beta}(\mathbf k,\mathbf q)}{\omega+\epsilon_{\mathbf k+\mathbf q,\beta} - \epsilon_{\mathbf k\alpha}+l\Omega+i\gamma_{\alpha\beta}},
	\end{align}
	where $\gamma_{\alpha\beta}=\gamma_{\mathbf k\alpha}+\gamma_{\mathbf k+\mathbf q,\beta}=2\Gamma$ for the reservoir, and the vertices, with the ordering matrices $\Oc_a$, are
	\begin{align*}
		&U^{a,lm}_{\beta\alpha}(\mathbf k,\mathbf q)=\sum_r\Big\langle\phi^{(r+l)}_{\mathbf k+\mathbf q,\beta}\Big|\Oc_a\Big|\phi^{(r+m)}_{\mathbf k\alpha}\Big\rangle,\\
		&V^{b,nl}_{\alpha\beta}(\mathbf k,\mathbf q)=\sum_s\Big\langle\phi^{(s+n)}_{\mathbf k\alpha}\Big|\Oc_b\Big|\phi^{(s+l)}_{\mathbf k+\mathbf q,\beta}\Big\rangle.
	\end{align*}
	For a Hermitian scalar vertex and $m=n=0$, $U^{l0}_{\beta\alpha}V^{0l}_{\alpha\beta}=|M^l_{\alpha\beta}|^2$ with $M^l_{\alpha\beta}=\sum_r\big\langle\phi^{(r+l)}_{\mathbf k+\mathbf q,\beta}\big|\Oc\big|\phi^{(r)}_{\mathbf k\alpha}\big\rangle$, so that
	\begin{equation*}
		\overline{\Pi}^R_{\mathbf q}(\omega)=\frac1N\sum_{\mathbf k,\alpha\beta,l}\frac{f_{\mathbf k\alpha}-f_{\mathbf k+\mathbf q,\beta}}{\omega+\epsilon_{\mathbf k+\mathbf q,\beta}-\epsilon_{\mathbf k\alpha}+l\Omega+i\gamma_{\alpha\beta}}\big|M^l_{\alpha\beta}\big|^2,
	\end{equation*}
	which is Eq.~\eqref{eq:pol} in the main text. We work in the limit $\gamma_{\alpha\beta} = 2\Gamma \to 0$, for vanishing system-bath coupling. Eq.~\eqref{eq: supppol} omits the change of the occupations caused by the fluctuation itself. In the normal state this change does not contribute at linear order at the onset, so this is the complete response function for  $\mathbf q \neq 0$. In an ordered state, this is the response function for $|\omega| \gg \gamma$.
	
	\subsubsection{Stoner criterion}
	Varying the action with respect to the quantum component and then putting $\delta m^q=0$ gives the retarded fluctuation equation
	\begin{equation} \label{eq: supplin}
		\sum_{b,n}\Big[(I^{-1})^{ab}_{mn}-\Pi^{R,ab}_{\mathbf q,mn}(\omega)\Big]\delta m^{(n)}_b=\sum_{b,n}\Pi^{R,ab}_{\mathbf q,mn}(\omega)h^{(n)}_b,
	\end{equation}
	where $h_b$ is a Zeeman field conjugate to $\hat O_b$. The response of the mean field within RPA is given by
	\begin{equation*}
		\chi^R_{\mathbf q}(\omega)=\frac{\delta m}{\delta h} = \big[I^{-1}-\Pi^R_{\mathbf q}(\omega)\big]^{-1} \Pi^R_{\mathbf q}(\omega),
	\end{equation*}
	which diverges when
	\begin{equation} \label{eq: suppsc}
		\det\big[\mathbb{I}-I\Pi^R_\qvec(0)\big]=0 ,
	\end{equation}
	which is the Stoner criterion of the driven system defined at $\omega \to 0$. Eq.~\eqref{eq: suppsc}
	reduces to the Stoner criterion mentioned in the main text as follows. For a single channel, the $m = n = 0$ element of this reduces to the cycle-averaged response in Eq.~\eqref{eq:pol}. Higher harmonics of this are either forbidden by symmetries or negligibly small (see Sec.~(\ref{ssec: trunc})  for details), so we keep this block only. The matrix reduces to a number and we have the criterion,
	\begin{equation*}
		I\Pi^R_{\qvec, 00}(0) = I\overline{\Pi}^R_{\qvec}(0) = 1,
	\end{equation*}
	as written in the main text.
	
	\subsection{Stability of mean-field solutions}
	
	In this section, we discuss the stability of self-consistent mean-field solutions of the driven-dissipative Hubbard model.
	\subsubsection{Linear stability of gap equation} We can write a solution of the gap equation with multiple possible orders as
	\begin{align*}
		\bm{m_0} = \mathcal{I} \bm{M}([\bm{m_0}]),
	\end{align*}
	where $\bm{m}$ contains all harmonic of all orders $\bm m = \{m_a^{(l)}\}$, and $\mathcal{I} = \mathrm{diag}(I_a)$ is a diagonal matrix whose elements are interaction terms for each order, and same for different harmonics within that order. Now, consider that we move slightly away from the solution by $\bm \delta_0$, then the next solution we get is 
	\begin{align*}
		&\bm{m_0} + \bm\delta_1 = \mathcal{I} \bm{M}([\bm{m_0}+\bm\delta_0])\nonumber\\
		&   \bm\delta_1 - \bm \delta_0 = \left(\mathcal I J(\bm m_0) - \mathbb{I}\right)\bm \delta_0 \equiv \dot{\bm \delta}
	\end{align*}
	in iterative sense, where $J(\bm m_0) = \partial\bm M/\partial\bm m_0$ is the Jacobian of the map. So, for stability of the solution, the `stability matrix' $S=\mathcal{I} J(\bm m_0)-\mathbb{I}$ should have all negative eigenvalues. For a single, static, order parameter, $M = m_0 \Phi(m_0)$, and this leads to the condition $\Phi'(m_0)<0$. This is a necessary but not sufficient condition for physical stability.
	
	\subsubsection{Equilibrium}
	
	Now turning to the free-energy picture in equilibrium, we can write the mean-field grand potential per site is
	\begin{equation*}
		F(\bm m) = \sum_a\frac{m_a^2}{2I_a} + \Omega_{\rm G}(\bm m), 
	\end{equation*}
	where $\Omega_{\rm G}= - (T / N) \ln{\rm Tr}e^{-(H_0- \sum_a m_a\hat{\mathcal{O}}_a-\mu N_e)/T}$. Since $\partial\Omega_{\rm G}/\partial m_a = -M_a$, we can show that 
	\begin{equation*}
		S = -\mathcal{I} \partial^2 F \big|_{\bm m_0}.
	\end{equation*}
	Consequently, stability would be given by $\partial^2 F \big|_{\bm m_0}$, which would mean that $\bm m_0$ are local minima of the free energy $F$. Again, for a single order parameter, $\partial^2 F \big|_{m_0} = I^{-1}-M'(m_0) = -m_0 \Phi'(m_0)$, and the stability condition is $\Phi'(m_0) < 0$. Thus for equilibrium, the linear stability of gap equation is also a sufficient condition for dynamical stability.
	
	\subsubsection{Out-of-equilibrium dynamical stability}
	
	Away from equilibrium, there is no free energy, and stability is determined by the fluctuation dynamics. Expanding the Keldysh action in Eq.~\eqref{eq: suppkel} about an ordered steady state,  gives  Eq.~\eqref{eq: supplin}, which we write as
	\begin{equation*}\label{eq: suppfluct}
		D(\omega)\,\delta m(\omega)=\Pi^R(\omega)h(\omega)+\xi(\omega),~ D(\omega)=I^{-1}-\Pi^R(\omega),
	\end{equation*}
	Here, $\Pi^R$ is the full retarded response in the ordered steady state and $\xi$ is a Gaussian noise of variance $O(1/N)$ coming from the Keldysh block. Since $\Pi^R(\omega)$ is analytic for ${\rm Im} \omega>0$, $\delta m(t)=\int\frac{d\omega}{2\pi}e^{-i\omega t}D(\omega)^{-1}[\Pi^R(\omega) h(\omega)+\xi(\omega)]$ is a bounded, causal response if and only if $\det D(\omega) \neq 0$ for given ${\rm Im} \omega>0$. A zero at $\omega_p = \omega_r + i\Gamma_p$ with $\Gamma_p > 0$ defines a mode $\delta m \propto e^{-i\omega_r t}e^{\Gamma_p t}$, which grows. Hence the solution is dynamically stable if and only if $\det D(\omega) \neq 0$ for ${\rm Im}\omega \geq 0$. In particular, on the positive imaginary axis $\omega = iy$, the kernel $\Pi^R(iy) = \int_0^\infty dt\,e^{-yt}\, \Pi^R(t)$ is real and vanishes as $y \to \infty$. This implies $\det D(iy) \to \det\mathcal I^{-1} > 0$ and thus $\det D(0) < 0$ implies a change in sign so $\det D(iy)$ vanishes for some $y > 0$. Hence, the steady state has a growing mode which makes it unstable. At equilibrium, $\Pi^R(0)=J$, so $D(0)= I^{-1}-J=- I^{-1}S$, which leads to the same criterion as before.
	
	\subsubsection{Stability of the first continuous order against other orders} 
	
	Consider that as $U$ increases, order~1 at wavevector $\mathbf q_1$ is the first to appear, continuously from zero. We ask if another order~2, at $\mathbf q_2$, can grow on top of it. Let $M_a(m_1,m_2)$ be the order parameters of the steady state with mean fields $m_1,m_2$. Both channels have the coupling $I=U/3$, and a solution is given by $m_a=IM_a$. In the normal state $M_a \simeq \Pi_am_a$, with $\Pi_a \equiv \overline{\Pi}^R_{\qvec_a}(0)$ the static response of channel $a$, hence only order $a$ would set in at $U_S(\mathbf q_a)=3/\Pi_a$, and $\Pi_1>\Pi_2$.
	
	With only the order-1 field present, an electron at $\mathbf k$ is scattered only to $\mathbf k\pm\mathbf q_1$, so the steady state has no contribution from $\mathbf q_2$. Thus, $M_2(m_1,0)=0$ for every $m_1$. Linearizing $m_a = IM_a$ about the order-1 solution $(\overline m_1, 0)$ gives a Jacobian matrix which is triangular. Its eigenvalues are $\lambda_1 = 1 + I\overline m_1\Phi_1'(\overline m_1)$, which tests the amplitude of order~1 and is below one along a continuous onset, and $\lambda_2 = I\,\partial M_2/\partial m_2$, which can check the mixture of order~2. As $\overline m_1\to0$, $\partial M_2/\partial m_2\to\Pi_2$ and $I\to1/\Pi_1$, so
	\begin{equation*}
		\lambda_2\to\frac{\Pi_2}{\Pi_1}=\frac{U_S(\mathbf q_1)}{U_S(\mathbf q_2)}<1 .
	\end{equation*}
	Then, $D(0)=I^{-1}(\mathbb I-IJ)$ is also triangular and $\lambda_2<1$ implies $D_{22}(0)>0$. If instead $\lambda_2>1$, then $\det D(0)<0$ and, by our previous arguments, the state would be unstable for any bath coupling. The first continuous order is therefore stable against every other order at its onset, and by continuity over a finite range of $U$ above it, while an order that is not the first is unstable at its own onset. In equilibrium the same statement follows from Landau theory.

	\subsection{Square-lattice Hubbard model: ordering and topology} \label{sec:hubbard}
	
	In this section we discuss the driven Hubbard model on the square lattice, its ordering and its
	topology. Consider the single-band Hubbard model,
	\begin{equation} \label{eq: supphubbard}
		H=\sum_{ij\sigma}h_{ij}(t)c^\dagger_{i\sigma}c_{j\sigma}
		+U\sum_i n_{i\uparrow}n_{i\downarrow}+H_{\rm res},
	\end{equation}
	where $h_{ij}(t)$ carries the drive through the Peierls substitution $\kvec\to\kvec+\bm A(t)$ and
	$H_{\rm res}$ is the reservoir coupling of Eq.~\eqref{eq: supphres}.  With the spin density
	$\svec_i=\sum_{\alpha\beta}c^\dagger_{i\alpha}\bm\sigma_{\alpha\beta}c_{i\beta}$, where $\bm\sigma$
	are the Pauli matrices in spin space, one has $\svec_i^2=3(n_i-2n_{i\uparrow}n_{i\downarrow})$, hence $Un_{i\uparrow}n_{i\downarrow}=\frac U2n_i-\frac U6\svec_i^2$. The first term is absorbed into the chemical potential, which is therefore measured from $U/2$ - for $w'=0$ half filling is at $\mu=0$.  The second term treats the three spin components alike.  Writing it as $-\tfrac I2\sum_i\svec_i^2$ with $I = U / 3$, we decouple it on the Keldysh contour $C$ with a vector Hubbard--Stratonovich field $\bm m_i(t)$,
	\begin{align*}
		&	\exp\left\{i\!\int_C\!dt\sum_i\frac I2\svec_i^2\right\}\nonumber\\
		&~~~	=\int\!\mathcal{D}[\bm m]\exp\left\{i\!\int_C\!dt\sum_i
		\left(-\frac{\bm m_i^2}{2I}+\bm m_i\cdot\svec_i\right)\right\},
	\end{align*}
	whose saddle point is $\bm m_i=I\langle\svec_i\rangle$. Here, the SU(2)-symmetric decoupling keeps all three spin components.
	
	\begin{figure*}[t]
		\centering
		\includegraphics[width=0.8\textwidth]{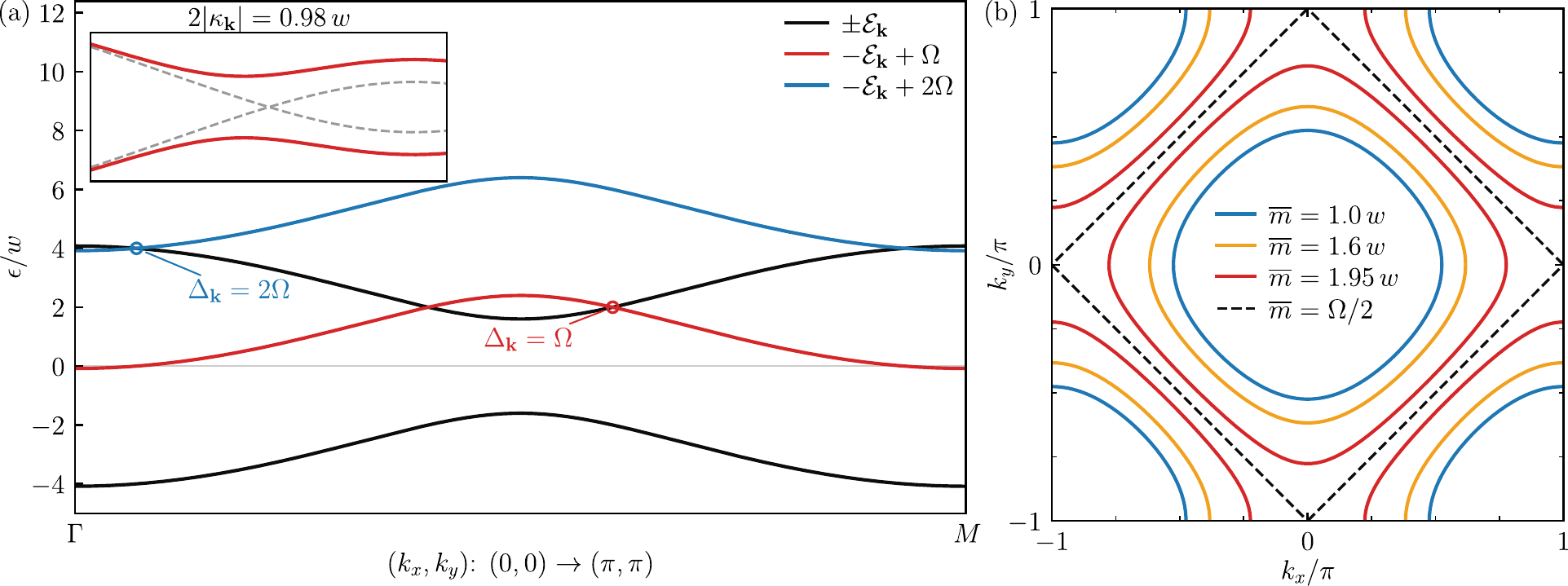}
		\caption{(a) Uncoupled replicas $\pm\mathcal E_{\mathbf k}$ (black), $-\mathcal E_{\mathbf k}+\Omega$ (red) and $-\mathcal E_{\mathbf k}+2\Omega$ (blue) along $(0,0)\to(\pi,\pi)$ at $\overline m=1.6w$. Circles mark the resonances $\Delta_{\mathbf k}=l\Omega$, Eq.~\eqref{eq:res}. Inset: exact quasienergies near the $p=1$ resonance (solid) and the uncoupled levels (dashed), with splitting $2|\kappa_{\mathbf k}|=0.98w$. (b) Resonant momenta of $p=1$ for three values of $\overline m$; dashed: the nesting diamond $\overline E_{\mathbf k}=0$, reached as $\overline m\to\Omega/2$. Drive frequency $\Omega=4w$ and amplitude $A_0=0.5$, with $w'=0$.}
		\label{fig:replicas}
	\end{figure*}
	
	\subsubsection{N\'eel order}  
	
	We take $\bm m_i(t)=m_{AF}(t)\hat ze^{i\Q\cdot\rvec_i}$ with $\Q=(\pi,\pi)$. In momentum space
	\begin{equation*}
		-\sum_i\bm m_i\cdot\svec_i = -m_{\rm AF}\sum^{\rm MBZ}_{\kvec\sigma}\sigma\psi^\dagger_{\kvec\sigma}\tau_1\psi_{\kvec\sigma},
	\end{equation*}
	which couples $\kvec$ to $\kvec+\Q$ and halves the zone. Here $\psi_{\kvec\sigma}=(c_{\kvec\sigma},c_{\kvec+\Q\sigma})^T$ with $\kvec$ in the magnetic zone (MBZ), the Pauli matrices $\tau_{1,2,3}$ acting on $(\kvec, \kvec+\Q)$, and $\sigma=\pm1$ labels the spin. In the four-component spinor of both spins the N\'eel vertex is $\tau_1\otimes\sigma_3$. At the saddle point $m_{\rm AF}=IM_{\rm AF}$, with the staggered moment $M_{\rm AF} = N^{-1}\sum_ie^{i\Q\cdot\rvec_i}\langle s^z_i\rangle$ of the $N$ sites.  When the mean-field synchronizes with the drive, the gap equation Eq.~\eqref{eq:gap} for each harmonic of $m_{\rm AF}(t)=\sum_l m_{\rm AF}^{(l)}e^{-il\Omega t}$, reads
	\begin{equation*}
		\frac{1}{I}=\frac{1}{Nm_{\rm AF}^{(l)}}\sum_{\kvec,\alpha}^{\rm MBZ}f_{\kvec\alpha}\sum_n
		\langle \phi^{(n)}_{\kvec\alpha} | \tau_1\otimes\sigma_3 | \phi^{(n+l)}_{\kvec\alpha} \rangle,
	\end{equation*}
	where $\alpha$ runs over the two bands and two spins at each $\kvec$.
	
	\subsubsection{Harmonic truncation}   \label{ssec: trunc}
	
	A circular drive rotates the vector potential by $\pi/2$ in a quarter period, $\bm A(t+T/4)=R_{\pi/2}\bm A(t)$, and a $C_4$ rotation about a site leaves the N\'eel pattern
	$e^{i\Q\cdot\rvec_i}$ unchanged. Hence $C_4$ combined with $t \to t + T / 4$ is a symmetry of the driven lattice and the reservoir, and it maps a solution with $m_{\rm AF}(t)$ onto one
	with $m_{\rm AF}(t+T/4)$. These solutions then obey $m_{\rm AF}^{(l)}e^{-il\pi/2}=m_{\rm AF}^{(l)}$, so $m_{\rm AF}^{(l)} = 0$ unless $l\equiv0\ ({\rm mod}\ 4)$. We keep $m_{\rm AF}^{(0)} \equiv \overline{m}$ only, and the first allowed correction $m_{\rm AF}^{(4)}$ is negligible at the amplitudes considered.
	
	\subsubsection{The mean-field Hamiltonian}
	
	For $w' = 0$ and static $m_{\rm AF}$, the problem is $2 \times 2$ at each $\kvec$ and spin, given by
	\begin{align} \label{eq: supphn}
		&	h_{\kvec\sigma}(t) = E_\kvec(t)\tau_3 - \sigma \overline{m} \tau_1, \\
		&	E_\kvec(t)=-2w\big[\cos(k_x+A_x(t))+\cos(k_y+A_y(t))\big] \nonumber.
	\end{align}
	whose cycle average is $\overline{E}_\kvec = -2wJ_0(A_0)(\cos k_x + \cos k_y)$. For real $\overline{m}$, $\tau_2 h_{\kvec\sigma} \tau_2 = -h_{\kvec\sigma}$, so $\mathcal{C} = \tau_2 K$ anticommutes with $h_{\kvec\sigma}-i\partial_t$. Each Floquet state at $\epsilon$ thus has a partner at $- \epsilon$ with reversed harmonic weights, and at $\mu = 0$ their occupations sum to one, so every chiral pair stays at half filling for any drive. Finite doping $\mu \neq 0$ or a next-nearest-neighbor hopping (for which $E_{\kvec+\Q} \neq -E_\kvec$) breaks this.
	
	\subsubsection{Resonances and the notch}
	
	Writing Eq.~\eqref{eq: supphn} in the Sambe basis, the harmonic-diagonal $\overline{E}_\kvec\tau_3-\sigma m_0\tau_1 - n\Omega$ give the replicas of the cycle-averaged bands,
	\begin{equation*}
		\epsilon_{\pm,n}=\pm\mathcal{E}_\kvec-n\Omega,
		~~
		\mathcal{E}_\kvec=\sqrt{\big(\overline{E}_\kvec\big)^2+m_0^2}.
	\end{equation*}
	and the $(n,n')$ block given by $E^{(n-n')}_\kvec\tau_3$ couples them. The replicas of opposite branches, $|e\rangle=|+_\kvec\rangle\otimes|n\rangle$ and $|o\rangle=|-_\kvec\rangle\otimes|n'\rangle$, are degenerate at the resonance $\Delta_\kvec\equiv 2\mathcal{E}_\kvec = p\Omega$ of Eq.~\eqref{eq:res} with $p = n - n'$, as shown in ~\Cref{fig:replicas}(a). Since the interbranch element of $\tau_3$ is $\overline{m} / \mathcal{E}_\kvec$, degenerate perturbation theory gives
	\begin{equation} \label{eq: suppheff}
		H_{\rm pair}=\begin{pmatrix}\epsilon_e&\kappa_\kvec\\ \kappa^*_\kvec&\epsilon_o\end{pmatrix},
	\end{equation}
	with $\delta_\kvec \equiv \epsilon_e - \epsilon_o = \Delta_\kvec - p\Omega$ and $|\kappa_\kvec| = \big|E^{(p)}_\kvec\big| \overline{m} / \mathcal{E}_\kvec$. At $\overline{m} = 0$, the drive is diagonal in the band basis and the replicas cross without hybridizing.
	The hybridization removes ordering weight. Writing Eq.~\eqref{eq:gap} at $l=0$ as $\Phi(\overline m)=(N\overline m)^{-1}\sum_{\mathbf k\alpha}f_{\mathbf k\alpha}v_{\mathbf k\alpha}$ with $v_{\mathbf k\alpha}=\sum_n\langle\phi^{(n)}_{\mathbf k\alpha}|\tau_1\otimes\sigma_3|\phi^{(n)}_{\mathbf k\alpha}\rangle$, each state contributes with weight $f_{\kvec\alpha} v_{\kvec\alpha}$. An uncoupled replica has $v_e = -\overline m/\mathcal E_{\mathbf k} = -v_o$ and a single harmonic, so its occupation is one Fermi factor of the band energy, $\bar f_e = n_F(\mathcal E_{\mathbf k} - \mu)$ and $\bar f_o = n_F(-\mathcal E_{\mathbf k} - \mu)$, which at $\mu=0$ are
	$n_F(\pm\mathcal{E}_\kvec)$. The two members of a resonant pair sit at the same quasienergy but are occupied differently, $\bar f_o - \bar f_e = \tanh(\mathcal E_{\mathbf k}/2T)$ at $\mu = 0$. Hybridizing them costs a factor $2q_\kvec - 1$ twice, where $|1\rangle=\sqrt{q_\kvec}|e\rangle+\sqrt{1-q_\kvec}|o\rangle$ and $|2\rangle=\sqrt{1-q_\kvec}|e\rangle-\sqrt{q_\kvec}|o\rangle$ diagonalize Eq.~\eqref{eq: suppheff} with $2q_\kvec-1=\delta_\kvec/\sqrt{\delta_\kvec^2+4|\kappa_\kvec|^2}$. One factor is for the vertex, as the two states lie in different harmonic sectors and $v_{\kvec\alpha}$ is diagonal in that index, so cross terms drop and $v_{1,2}=\pm(2q_\kvec-1)v_e$. The other factor is from the occupations, as the harmonic weights add and $\epsilon_1=\epsilon_2$ at the crossing, so $f_1 - f_2 = (2q_\kvec-1)(\bar f_e-\bar f_o)$. Hence
	\begin{equation*}
		f_1v_1+f_2v_2=\big(\bar f_o-\bar f_e\big)\frac{\overline{m}}{\mathcal{E}_\kvec}P_\kvec,
	\end{equation*}
	with $P_\kvec=(2q_\kvec-1)^2 = \delta_\kvec^2 / (\delta_\kvec^2+4|\kappa_\kvec|^2)$. Far from resonance $P_{\mathbf k}\simeq1$ and the pair contributes as it would without drive. At the crossing $P_{\mathbf k} = 0$, the pair is equally occupied and its contributions cancel. Hence
	\begin{align*}
		\Phi(\overline m) &\simeq \Phi_{\rm bg}(\overline m)-\delta\Phi_p(\overline m), \\
		\delta\Phi_p(\overline m) &= \frac1N\sum^{\rm res}_{\mathbf k\sigma}\frac{\bar f_o-\bar f_e}{\mathcal E_{\mathbf k}}\big(1-P_{\mathbf k}\big),
	\end{align*}
	where $\Phi_{\rm bg}(\overline m)$ is $\Phi$ without hybridization. The resonant momenta lie on the contour $\overline E_{\mathbf k}^{\,2}=(p\Omega/2)^2-\overline m^2$ as shown in ~\Cref{fig:replicas}(b). As $\overline m\to p\Omega/2$, it moves onto the nesting diamond $\overline E_{\mathbf k}=0$, the longest such contour and the one through the van Hove points, and above $p\Omega/2$ no momentum is resonant. The deficit therefore grows with $\overline m$ and vanishes abruptly at $p\Omega / 2$ - a notch in $\Phi(\overline m)$ with a sharp upper edge. Where the notch is deep enough, $\Phi=1/I$ has three solutions, as in the fold of Fig.~\ref{fig:meanf}.

	\begin{figure}[t]
		\centering
		\includegraphics[width=0.9\columnwidth]{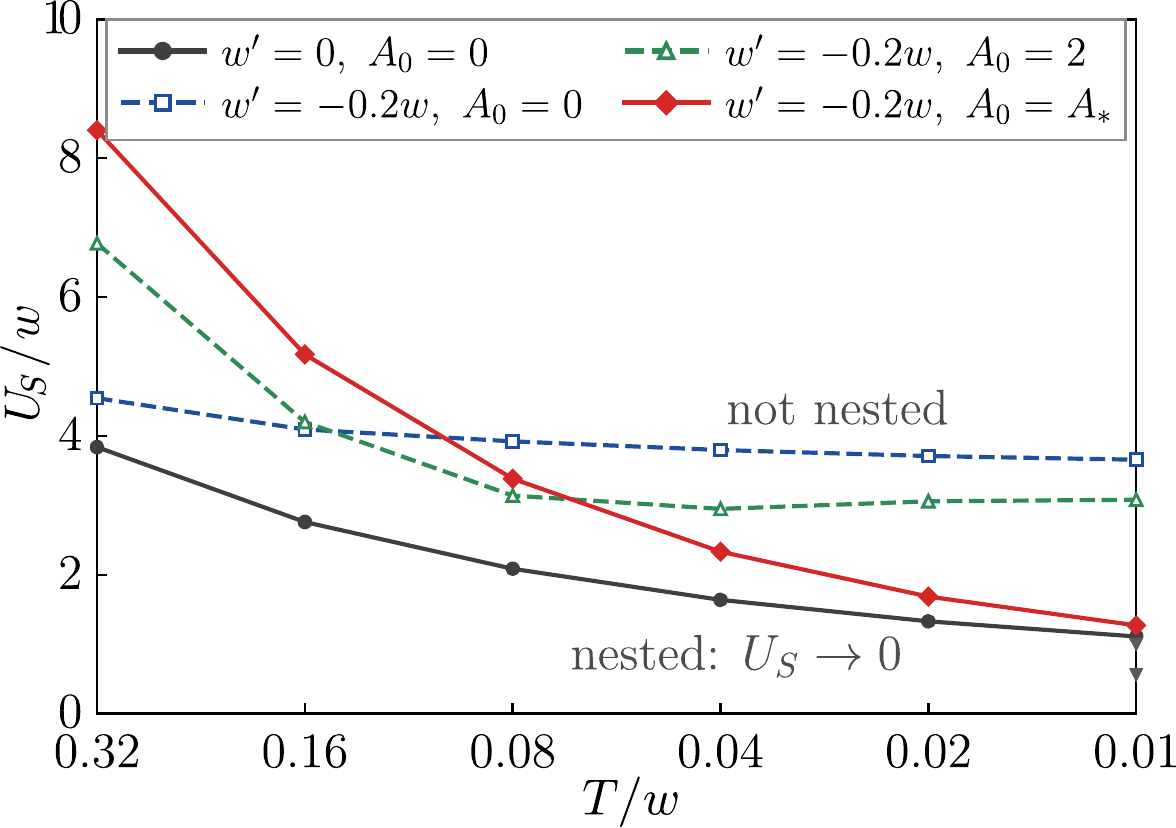}
		\caption{Normal-state Stoner threshold $U_S = 3/\overline{\Pi}_{\mathbf Q}(0)$ versus temperature, for $w'=0$ at $A_0=0$ (black) and for $w'=-0.2w$ at $A_0=0$ (blue), $A_0=2$ (green) and $A_0=A_*$ (red). Nested quasienergy bands make $\overline{\Pi}_{\mathbf Q}(0)$ diverge as $T \to 0$, so that $U_S$ vanishes in both cases. Away from $A_*$, the threshold saturates. Drive frequency $\Omega=4w$, $\mu=0$, linewidth $\gamma = T/20$, and an $800\times800$ momentum grid.}
		\label{fig:nesting}
	\end{figure}
	
	\subsubsection{Winding numbers}
	When $\Omega$ is below the bandwidth, the topology is described by winding numbers, each of which
	belongs to a gap rather than to a band.  Let
	$U_\kvec(t)=\mathcal{T}\exp\big[-i\!\int_0^t\!dt'h_{\kvec\sigma}(t')\big]$.
	Since $U_\kvec(T)\neq\mathbb{I}$ in general, one first deforms it into a loop:
	with $H^\epsilon_{\rm eff}(\kvec)=(i/T)\log U_\kvec(T)$, the branch cut of the
	logarithm taken at quasienergy $\epsilon$,
	\begin{equation}
		U_\epsilon(\kvec,t)=U_\kvec(t)e^{iH^\epsilon_{\rm eff}(\kvec)t}
		\label{eq:Uper}
	\end{equation}
	satisfies $U_\epsilon(\kvec,0)=U_\epsilon(\kvec,T)=\mathbb{I}$ and maps the
	three-torus $(k_x,k_y,t)$ into $U(2)$.  Such maps carry an integer winding number
	\cite{Rudner2013},
	\begin{equation}
		W_\epsilon=\frac{1}{8\pi^2}\!\int\!dtd^2k\;
		{\rm Tr}\big(L_t\big[L_{k_x},L_{k_y}\big]\big),
		~~
		L_\mu=U_\epsilon^{-1}\partial_\mu U_\epsilon .
		\label{eq:Wdef}
	\end{equation}
	The gaps at $\epsilon=0$ and $\Omega/2$ give $W_0$ and $W_\pi$.  The Chern number of the band
	between $\epsilon=-\Omega/2$ and $0$ is $C=W_0-W_\pi$ (the band between $0$ and $\Omega/2$ has
	$-C$), and $|W_\epsilon|$ is the number of chiral edge modes crossing the gap at $\epsilon$ in a
	ribbon.

	\begin{figure}[t]
		\centering
		\includegraphics[width=0.45\textwidth]{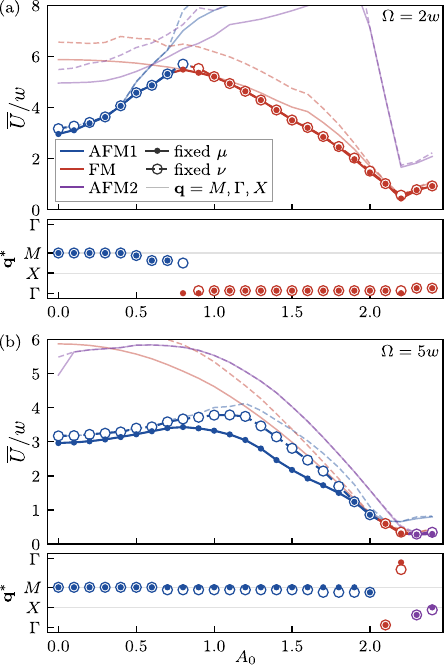}
		\caption{Leading magnetic instabilities at the parameters of Ref.~\cite{Herre2026} for (a) $\Omega=2w$ and (b) $\Omega=5w$. Top: onset $\overline{U}$ of the order with the lowest onset over $\mathbf q$ on the path $\Gamma \to X \to M \to\Gamma$, at fixed $\mu$ (solid) and fixed $\nu$ (dashed), colored by the nearest symmetry point - AFM1 ($M$), FM ($\Gamma$) and AFM2 ($X$). Thinner lines show $\mathbf q$ held at $M$, $\Gamma$ or $X$. Bottom: position of the leading $\mathbf q = \mathbf{q}^*$ on the path, at fixed $\mu$ (filled) and fixed $\nu$ (open). Here, $w'=-0.2w$, $\nu=1.2$ in the normal state, $I = U/2$ and temperature $T=0.01w$, on a $128 \times 128$ momentum grid.}
		\label{fig:herre}
	\end{figure}

	\subsubsection{Dynamical nesting}
	\label{sec:nesting}
	
	Nesting, $E_{\kvec+\Q}=-E_\kvec$, gives antiferromagnetic ordering at
	arbitrarily weak $U$ at half filling.  A next-nearest-neighbor hopping $w'$
	breaks it,
	\begin{align}
		E_\kvec(t)&=-2w\big[\cos(k_x{+}A_x)+\cos(k_y{+}A_y)\big]\nonumber\\
		&~~		-4w'\cos(k_x{+}A_x)\cos(k_y{+}A_y),
		\label{eq:disp}
	\end{align}
	and in the magnetic zone
	$h_{\kvec\sigma}(t)=\xi^+_\kvec\tau_0+\xi^-_\kvec\tau_3-\sigma \overline{m}\tau_1$ with
	$\xi^\pm_\kvec=\tfrac12(E_\kvec \pm E_{\kvec+\Q})$. The whole of $w'$ is only contained in $\xi^+_\kvec=-4w'\cos(k_x{+}A_x)\cos(k_y{+}A_y)$ multiplying $\tau_0$.  Being a
	$c$-number, $\xi^+_\kvec$ commutes with the rest of $h_{\kvec\sigma}(t)$ at all times, so
	it enters the quasienergies only through its cycle average,
	\begin{equation}
		\langle\xi^+_\kvec\rangle_t=-4w'J_0(\sqrt2A_0)\cos k_x\cos k_y .
		\label{eq:defect}
	\end{equation}
	This vanishes at the amplitude $A_*=2.4048/\sqrt2$ and the quasienergy spectrum becomes that of a perfectly nested lattice although the instantaneous bands are not nested.  
	
	In \Cref{fig:nesting} we plot the Stoner threshold in the normal state where the quasienergy bands are nested, as $T\to0$ and the threshold vanishes.
	
	\subsection{Fixed filling vs fixed chemical potential}
	\label{sec:herre}
	This follows the same model of the work by Herre \textit{et al.}~\cite{Herre2026} on the magnetic instabilities of the circularly driven square lattice at $w' = -0.2w$ and filling $\nu = 1.2$, for $\Omega = 2w$ and $5w$, where order was computed using Stoner criterion from the normal state Floquet RPA susceptibility with effective interaction $I = U/2$ and linewidth $\gamma = 0.1w$. %Each instability is labelled by the high-symmetry point nearest to the maximum of the susceptibility - AFM1 near $M = (\pi, \pi)$, FM near $\Gamma = (0, 0)$ and AFM2 near $X = (\pi, 0)$. 
	In ~\Cref{fig:herre} we revisit this parameter regime using $\Phi(\overline{m})$, which can also capture discontinuous onsets. We calculate the critical interaction for onset of order as $\overline{U}(\qvec) = 2 / \max_{\overline{m}}\Phi_{\mathbf q}$ (with $\gamma\to 0$), including the commensurate orders defined at $\mathbf{q} = M$, $\Gamma$ and $X$. The order that sets in first has the lowest $\overline{U}$ over $\mathbf{q}$ on the path $\Gamma \to X \to M \to \Gamma$, and it is named accordingly by the nearest symmetry point. Incommensurate orders are labeled as `AFM1' when $\mathbf{q}$ is near $M = (\pi, \pi)$, `FM' when $\mathbf{q}$ is near $\Gamma = (0, 0)$ and `AFM2' when $\mathbf{q}$ is near $X = (\pi, 0)$.
	We do it, once at fixed $\mu$, where the reservoir chemical potential is set by the filling of the normal state $\nu = 1.2$, as well as at fixed $\nu$, where the chemical potential is changed in the ordered state for each $\overline{m}$. At $\Omega = 2w$, AFM1 sets in first up to $A_0 \simeq 0.85$ and FM beyond, which resembles the AFM1 to FM transition in Ref.~\cite{Herre2026}. At $\Omega = 5w$, AFM1 transitions to FM around $A_0 \simeq 1.9$, which then goes into AFM2. Here, the drive moves the saddles to $\epsilon_X = 4w'J_0(\sqrt2A_0)$, at a distance $\delta = \mu-\epsilon_X$ below $\mu$ that shrinks with $A_0$. In the FM and AFM2 regimes, the leading $\qvec$ lies along $\Gamma \to X$ or $M \to  \Gamma$ and is generally incommensurate. The fixed $\mu$ and fixed $\nu$ cases largely differ in the AFM1 regime at $\Omega = 5w$. At fixed $\mu$ the leading order is the commensurate N\'eel order up to $A_0 \simeq 1.9 $, and its onset is discontinuous. As in the main text, the N\'eel order splits the filled saddle states to $\epsilon_X \pm \overline m$, which contribute to $\Phi_{\mathbf{Q}}$ only once the higher energy level rises above $\mu$, at $\overline m > \delta$. At fixed filling, $\mu$ readjusts as the levels split and the N\'eel onset moves to larger $U$. As a result, an incommensurate spiral order sets in first.

\end{document}